# CONSTRAINTS ON SELF-INTERACTING DARK MATTER

Andrew A. de Laix and Robert J. Scherrer
*Department of Physics, The Ohio State University*
*Columbus, Ohio 43210*

Robert K. Schaefer
*Bartol Research Institute, University of Delaware*
*Newark, Delaware 19716*

## Abstract

We consider the growth of density perturbations in the presence of self–interacting dark matter, SIDM, proposed by Carlson, Machacek and Hall (1992). We determine the range of values for the coupling constant $\lambda$ and the particle mass $m'$, for which the power spectrum lies in the "allowed" range based on constraints from the IRAS galaxy survey and damped Lyman–$\alpha$ systems. Our results show that no combination of parameters can meet both limits. We consider constraints on the $2 \leftrightarrow 2$ scatterings which keep the SIDM particles in pressure equilibrium, and we show that if such interactions maintain pressure equilibrium down to the present, they will be strong enough to disrupt galaxy mergers and may lead to stripping of galaxy halos as galaxies move through the dark matter background of these particles. Hence, we also investigate the evolution of large-scale structure in the SIDM model when the particles drop out of pressure equilibrium at some higher redshift. The resulting free-streaming leads to an additional suppression of small-scale perturbations, but it does not significantly affect our results.

# 1 Introduction

Most models for large-scale structure have assumed two possibilities for the dark matter in the universe: hot dark matter, which is relativistic at the time the horizon is galaxy-sized, or cold dark matter, which is nonrelativistic. Because of free-streaming, hot dark matter has a power spectrum which falls off sharply at small scales, resulting in well-known problems with early galaxy formation (White, Frenk, and Davis 1983; White, Davis, and Frenk 1984; Schaeffer and Silk, 1988). Cold dark matter has more small-scale power, but with the COBE normalization it may have too much power on small scales (see, for example, Schaefer and Shafi 1994 and references therein). This has led to renewed interest in models with a power spectrum intermediate between these two cases, such as a mixture of hot and cold dark matter (see, for example, Schaefer and Shafi, 1993; Klypin *et al.* 1993). Another possibility has been suggested by Carlson, Machacek and Hall (CMH) (1992). In their model, called self–interacting dark matter (SIDM), the dark matter particle interacts strongly with itself but weakly with ordinary matter. Thus, SIDM can have number–changing interactions which conserve entropy when the temperature has fallen below the mass of the particle. The result is that the particles convert their rest energy into kinetic energy, heating themselves to a higher temperature than ordinary nonrelativistic particles. CMH suggested that this might lead to a power spectrum intermediate between hot and cold dark matter, and this possibility was explored further by Machacek (1994).

In this paper we examine perturbation growth in the SIDM model in more detail. When the SIDM particle is in the self-interacting regime, pressure suppresses perturbation growth; this has been modeled by CMH and by Machacek (1994), who assumed that $2 \leftrightarrow 2$ scattering reactions would keep the particle in pressure equilibrium down to the present. However, there are cosmological constraints on any dark matter particle in pressure equilibrium at present. In particular, the scattering cross section for such a particle is so large that $2 \leftrightarrow 2$ scatterings between dark matter particles in the halos of colliding galaxies will significantly alter the process of galaxy merger. Furthermore, a galaxy halo moving through the background sea of SIDM particles will experience ram pressure which can strip the halo off of the baryonic component of the galaxy. Hence, it is important to consider the alternative possibility that the $2 \leftrightarrow 2$ interactions decouple at some higher redshift. In this case, the SIDM particles can free-stream out of overdense regions, erasing previously-existing perturbations, and reducing further the amplitude of fluctuations on small scales.

Thus, we are motivated to examine the SIDM model for the case where these scattering interactions decouple at some higher redshift. In this case, the effects of free-streaming must also be included in the calculation.

In the next section, we discuss the general features of the self–interacting model. In §3, we assume that pressure equilibrium is maintained down to the present and repeat the calculation of Machacek (1994) for a wide range of SIDM parameters. We calculate the range of values of the SIDM mass and coupling constant which gives an acceptable power spectrum. In §4, we derive cosmological constraints on the $2 \leftrightarrow 2$ scattering interactions and discuss the effects of free-streaming if these interactions decouple. Our conclusions are summarized in §5. We find that in general we cannot produce enough small-scale power to account for damped Ly$\alpha$ systems without producing too much power on cluster progenitor ($8h^{-1}$ Mpc) scales.

## 2   The Self–Interacting Model

The properties of self-interacting dark matter are discussed in detail by CMH; here we summarize briefly the results which we need for our calculations. The evolution of SIDM is determined by the temperature at which the SIDM becomes non-relativistic, $T_{nr}$, and the temperature at which the rate for the number-changing interactions drops below the expansion rate and these number-changing reactions decouple, $T_d$. (The SIDM particle is assumed to decouple from the thermal background at a temperature much higher than either $T_{nr}$ or $T_d$). The evolution of SIDM is interesting only if $T_{nr} > T_d$, i.e., the SIDM particle becomes non-relativistic before the number-changing interactions decouple. In this case, the evolution of SIDM can be divided into three regimes: the relativistic regime ($T > T_{nr}$), the self–interacting regime ($T_{nr} > T > T_d$), and the non–relativistic regime ($T_d > T$). In the relativistic regime, the temperature is greater than the mass of the particle, and SIDM behaves exactly like ordinary relativistic matter; the number-changing interactions have no effect on the particle abundance or temperature, so the energy density scales as $a^{-4}$ and the temperature as $a^{-1}$. In the non-relativistic regime, the particle is decoupled and non-relativistic, so it behaves like any other non-relativistic particle out of thermal equilibrium, i. e. $T \propto a^{-2}$ and

$\rho \propto a^{-3}$.

If $T_d > T_{nr}$, the particle moves directly from the relativistic regime to the non-relativistic regime, so there is nothing particularly interesting about its evolution. However, in the opposite case, there is a regime (the self–interacting regime), where the particle is non-relativistic and out of equilibrium with the thermal background, but number-changing self-interactions are still important: this is the regime in which the evolution of SIDM differs from either relativistic or nonrelativistic matter. In this regime the energy and entropy densities for SIDM with mass $m'$, temperature $T'$, and spin degeneracy $g'$ are given by a distribution with zero chemical potential:

$$\rho = g'm' \left(\frac{m'T'}{2\pi}\right)^{3/2} e^{-m'/T'} \qquad (1)$$

and

$$s = \frac{\rho'}{T'}. \qquad (2)$$

Primed variables will always refer to the SIDM. (These results and those which follow are taken from CMH). Since the the entropy per comoving volume is conserved, the product $sa^3$ is a constant. One may infer from equations (1) and (2) that the temperature, $T'$, is approximately proportional to the inverse of the logarithm of the scale factor,

$$T' \sim \frac{1}{\log(a)}, \qquad (3)$$

in the self–interacting regime. To conserve entropy, the gas annihilates some of its number to reheat itself, thus avoiding the usual power law cooling, i.e., $1/a$ for a relativistic gas and $1/a^2$ for a nonrelativistic gas. It is also interesting to note that the energy density falls slightly more rapidly than for nonrelativistic matter which cannot change its number density per comoving volume,

$$\rho' \sim \frac{1}{a^3 \log(a)}, \qquad (4)$$

an intermediate behavior between relativistic and non–relativistic particles. Once the number changing interactions are no longer sufficient to keep up with the universal expansion, the particle has entered the non-relativistic epoch.

As has already been mentioned, one expects that the interactions which allow number changing interactions will eventually become so weak that they will no longer be able to keep up with the universal expansion. To remain self–interacting, the number changing interactions must be fast enough to maintain a zero chemical potential distribution. If $\Gamma$ is the rate for a particular interaction and $N'$ is the number of particles per comoving volume, CMH infer the condition for self interaction

$$\Gamma \geq \frac{\dot{N'}}{N'}. \qquad (5)$$

When this inequality no longer holds, the SIDM is decoupled and behaves exactly like an ordinary nonrelativistic relic. The temperature of the SIDM at decoupling is defined as $T'_d$.

CMH motivated their discussion with a "toy" scalar model with $3 \to 2$ interactions resulting from a $\phi^5$ term in the Lagrangian. It is unnecessary to consider a particular model since, for any particle with $3 \to 2$ interactions, one may write the rate as

$$\Gamma(3 \to 2) = \lambda m' \left(\frac{n'}{m'^3}\right)^2, \qquad (6)$$

where $\lambda$ is a dimensionless quantity related to the coupling, $n'$ is the number density of the particles, and $m'$ is the mass. The quantity $\lambda$ is model-dependent and can depend on $T'$, so in the expressions below it is implicitly evaluated at the decoupling temperature $T'_d$.

Using the decoupling condition along with the expression for the rate, CMH have calculated an approximate relation between the mass and the decoupling temperature for both the early and late decoupling. For early decoupling, the universal expansion is dominated by radiation at $T'_d$, while for late decoupling the expansion is dominated by the SIDM. For the latter case, CMH have shown that

$$\frac{m'}{T'_d} + \frac{3}{2} \ln\left(\frac{m'}{T'_d}\right) = \frac{2}{3} \ln\left(\frac{\lambda'}{\xi' \Omega_0 h^2}\right) + 38.06, \qquad (7)$$

where $\Omega_0$ is the present energy density (assumed to be dominated by the SIDM) in units of the critical density, and $h$ is the Hubble parameter in units of 100 km s$^{-1}$ Mpc$^{-1}$. The quantity $\xi$ is the ratio of entropy in radiation, $s$, to entropy in SIDM, $s'$, and $\xi'$ is related to $\xi$ by

$$\xi' = \xi g'^{1/4} \tag{8}$$

where $g'$ is the number of spin degrees of freedom of the SIDM particle. Also, $\lambda'$ is related to $\lambda$ by

$$\lambda' = \lambda g'^{7/4}. \tag{9}$$

If the universe is dominated by radiation at decoupling then the relationship between the mass and decoupling temperature takes a different form:

$$\frac{m'}{T'_d} + 2\ln\left(\frac{m'}{T'_d}\right) = \frac{3}{4}\ln\frac{\lambda'}{\Omega_0 h^2} - \frac{5}{4}\ln\xi' + 43.39. \tag{10}$$

For both equations, the value of $\xi$ is not a free parameter but a constant related to the decoupling temperature $T'_d$ by

$$\frac{\rho'_{now}}{s_{now}} = \frac{T'_d}{\xi}, \tag{11}$$

where we have used the fact that the energy density of the SIDM and the entropy of the radiation both have the same dependence on the scale factor after decoupling. Evaluating this expression one gets

$$T'_d = (3.6 \text{ eV})\Omega_0 h^2 \xi. \tag{12}$$

Now it is clear from eqs. (7) and (12) that when the SIDM decouples while dominating the expansion of the universe, the decoupling temperature is independent of $\Omega_0 h^2$. A lower limit on the value of $\xi$ can be derived from the nucleosynthesis constraint on the number of extra degrees of freedom. For $\Delta N < .3$ (Walker et al. 1991), CMH obtain $\xi > 17$. With eq. (12) we now have sufficient information to relate the mass to the decoupling temperature for a given $\lambda'$, but because of the logarithmic dependence on the coupling constant, the result for the decoupling temperature is roughly model independent in so far as $3 \to 2$ processes are the dominant number–changing interactions. CMH also considered a pseudoscalar with a dominant channel of $4 \to 2$ which produced freeze out ratios of $T'/m'$ about twice that of the scalar case. We have ignored this second case as it will not significantly change our conclusions.

# 3  Power Spectra Calculations

In this section, we extend the calculations of Machacek (1994) who assumed that $2 \leftrightarrow 2$ scattering interactions maintain pressure equilibrium down to the present. In the next section, we will relax this assumption and allow for the possibility of free-streaming. To conduct detailed simulations of the growth of SIDM fluctuations, one needs to consider general relativistic perturbation theory. Here we treat the SIDM like an ideal fluid; in the next section we will consider the case of collisionless particles. Many equivalent constructions exist; however, the most convenient choices are those formulations which are gauge invariant and thus avoid superfluous gauge modes and other ambiguities. We will use the formulations derived by Kodama and Sasaki (1984) in this work.

If pressure equilibrium is maintained down to the present, we use the perfect fluid equations appropriate for particles in thermal contact. Kodama and Sasaki have constructed a set of gauge invariant variables which will reduce to the following in the Newtonian limit in Fourier space: $\Delta$, the density perturbation; $V$, the velocity perturbation; $\Pi$, the anisotropic pressure perturbation and $\Gamma$, the entropy perturbation. For the $i^{th}$ component of a multiple component fluid, the evolution in conformal time $\tau$ ($dt = ad\tau$), of the Fourier components $\Delta_{ci}(\tau, k)$ and $V_i(\tau, k)$ is given by

$$\dot{\Delta}_{ci} + 3\frac{\dot{a}}{a}(c_i^2 - w_i)\Delta_{ci} = -3\frac{\dot{a}}{a}w_i\Gamma_i - k(1+w_i)V_i + 3\frac{\dot{a}}{a}\frac{1+w_i}{1+w}\left[c_s^2\Delta + w\Gamma - \frac{2}{3}w\Pi\right] \tag{13}$$

and

$$\dot{V}_i + \frac{\dot{a}}{a}V_i = 3\frac{\dot{a}}{a}c_i^2(V_i - V) - \frac{3}{2k}\left(\frac{\dot{a}}{a}\right)^2(\Delta + 2w\Pi) + \frac{k}{1+w_i}\left[c_i^2\Delta_{ci} + w_i\Gamma_i - \frac{2}{3}w_i\Pi_i\right] \tag{14}$$

where the dot denotes the derivative with respect to the conformal time, $c_s^2 = (\partial p/\partial \rho)$ is the sound speed, and $w$ is the ratio of total pressure to total energy. $\Delta$, $V$, $\Pi$ and $\Gamma$ are the total perturbations summed over all components:

$$\Delta = \rho^{-1}\sum \rho_i \Delta_{ci} \tag{15}$$

$$V = (\rho + p)^{-1}\sum (\rho_i + p_i)V_i \tag{16}$$

$$\Pi = p^{-1} \sum p_i \Pi_i \tag{17}$$

$$\Gamma = p^{-1} \sum \left[ p_i \Gamma_i + (c_i^2 - c_s) \rho_i \Delta_{ci} \right]. \tag{18}$$

Subscripts refer to the components; variables without subscripts refer to the total. For the SIDM model in pressure equilibrium, $\Gamma$ and $\Pi$ are negligible.

Using these equations for a fluid in pressure equilibrium, we have repeated the calculation of Machacek (1994) for a wide range of values for $\lambda$ and $m'$, with $\Omega = 1$ and $h = 1/2$. Three massless neutrinos were included along with the photons, but the baryons were neglected. For simplicity, the photons and neutrinos were treated as perfect fluids, ignoring free–streaming; this is adequate when the growth of SIDM perturbations is the primary concern.

To constrain SIDM models, we have used two limits on the linear mass fluctuations determined by Schaefer and Shafi (1994). The filtered rms mass function is defined by

$$\sigma^2 = \frac{1}{2\pi^2} \int P(k) |W(k)|^2 k^2 dk, \tag{19}$$

where $P(k)$ is the power spectrum and $W(k)$ is the Fourier transform of the window function. The first such constraint is an upper limit on the amplitude of mass fluctuations based on data from the IRAS (InfraRed Astronomy Satellite) galaxy survey (Schaefer and Shafi 1994). For a spherical top–hat filter with a radius of $8h^{-1}$ Mpc, Schaefer and Shafi (1994) estimate that the mass fluctuation should be less than 0.8. This is in good agreement with the constraint for not overproducing x-ray clusters (White, Efstathiou, & Frenk, 1993), as $8h^{-1}$ Mpc is roughly the size of an overdense region which forms a cluster of galaxies. However, both the estimate from the IRAS survey and the x-ray clusters were derived using the "spherical collapse model" and may be systematically in error, so to be conservative here we have chosen a looser bound $\sigma(8h^{-1}$ Mpc$) \lesssim 1$.

Our second constraint is derived from the observation of damped Ly$\alpha$ systems (Klypin *et al.* 1995) which is roughly $\sigma(.13h^{-1}$ Mpc$) \gtrsim 3.0$. This constraint can be derived as follows. First of all, it is highly likely that damped Lyman–$\alpha$ systems correspond to systems with mass $> 10^{10}$ $M_\odot$ if they are to produce the observed neutral hydrogen column densities using $\Omega_{baryon} = 0.05$, consistent with nucleosynthesis (Walker, *et al.* 1991). This corresponds to a Gaussian filter radius of 0.13 $h^{-1}$ Mpc. One can imagine interesting ways to utilize smaller dark matter halos, but this will not change our argument, since going to smaller scales in interesting SIDM models will not significantly change the amplitude of mass fluctuations. Second, the most difficult constraint to satisfy is that the fraction of the mass in neutral hydrogen gas observed in these objects at $z = 3.2$ is $\Omega_{gas} = 0.0058 \pm 0.0011\ (0.5/h)$ (Lanzetta, et al. 1995). We next assume that all of the baryons remain as neutral hydrogen even after collapse. This is unrealistic, but if the models cannot satisfy this constraint, they will be clearly not be able to satisfy a more realistic constraint. If we require that the models only need to satisfy the *two-$\sigma$ lower bound*, i.e., $\Omega_{gas} \geq 0.0034\ (0.5/h)$, this implies that $\sigma(0.13\ h^{-1}\ \text{Mpc}, z = 3.2) \geq 3.0$ using the Press-Schecter formulas applied as described in Klypin *et al.* (1995).

With these two constraints it is possible to bracket the parameters of the allowable SIDM models. Fig. 1 shows the values of $m'$ and $\lambda$ which violate each constraint. Our power spectra were normalized to the latest value for the COBE quadropole, $Q = 20\mu K$ (Bunn, Scott, and White, 1995; Tegmark and Bunn 1995 and references therein). Masses above the solid curve produce too large a fluctuation on $8h^{-1}$ Mpc scales. The excluded region is marked with horizontal hatching. Masses below the dot–dashed curve do not produce sufficient damped Ly$\alpha$ systems. Here the excluded region is marked with vertical lines. No combination of $m'$ and $\lambda$ can simultaneously satisfy both criteria. The reason for this is apparent in Fig. 2. This figure shows the transfer functions, solid lines, for three different SIDM masses (1, 10 and 100 KeV) with $\lambda = 10^{-3}$. The transfer function describes the relative growth of perturbation amplitudes as a function of (inverse) scale. Also shown are the HDM and CDM transfer functions (Bardeen *et al.* (1986)) as well as the mixed (20% HDM and 80% CDM) dark matter transfer function (Schaefer and de Laix 1995). The sharp turnover in the SIDM transfer function caused by the Jeans suppression kills the small-scale structure. To meet the Ly$\alpha$ constraint, the turnover point of the transfer function has to be pushed to $k$ values large enough to produce excessive large scale structure. The mixed dark matter model is the only one to satisfy both constraints. The loss of some of the large scale structure due to the streaming of the HDM along with the preservation of small-scale power by the CDM allows both constraints to be met simultaneously.

The route to fixing the ills of the standard CDM picture is clear from the constraints imposed here. One must reduce amplitude of fluctuations on the tens of Mpc scale taking care not to remove too much smaller-scale power. If one wants to reduce the growth of large scale power ($\geq 8h^{-1}$ Mpc) with free streaming (or non-negligible pressure) of relic particles, one must also add a separate source of power on smaller scales to account for the early formation of structure. Damping the growth of density fluctuations in a single relic dark matter particle will not do the whole job.

## 4 Constraints on $2 \leftrightarrow 2$ Scattering

In the previous section, we followed CMH and Machacek (1994) in assuming that $2 \leftrightarrow 2$ scatterings would maintain pressure equilibrium down to the present. In this section, we consider constraints on such interactions. Let $\bar{n}$, $\sigma$, and $v$ be the mean number density, $2 \leftrightarrow 2$ scattering cross sections, and velocity of the SIDM particles respectively. The condition that pressure equilibrium be maintained down to the present is equivalent to the condition $n\sigma v > H_0$ at present. As noted by CMH, the SIDM particles cannot have higher velocities than the current galaxy rotation velocities, in order for them to be bound into the galaxy halos, which gives $v \lesssim 10^{-3}$. Taking $1/H_0 = 3000h^{-1}$ Mpc, we obtain an upper bound on $1/\bar{n}\sigma$, the mean free path for $2 \leftrightarrow 2$ scattering:

$$1/\bar{n}\sigma \lesssim 3h^{-1} \text{ Mpc}. \tag{20}$$

However a stronger constraint can be obtained if we consider the condition required for the galaxy to form in the first place. In order for a galactic density perturbation to become Jeans' unstable (and collapse), the pressure $p \sim m'\bar{n}\langle v^2 \rangle$ must not be so large so that the perturbations just undergo acoustic oscillations. The cosmological Jeans' instability condition can be expressed as follows (Abbott and Wise, 1984). To make galaxies, the "sound" horizon today must be smaller than the the size of the galaxy:

$$c_s/H_0 \lesssim 100 \ h^{-1} \text{ kpc}, \tag{21}$$

Since $v \approx c_s$, we can combine eq. (21) with the condition for pressure equilibrium to get the limit

$$1/\bar{n}\sigma \lesssim 100 \ h^{-1} \text{ kpc}. \tag{22}$$

If we assume then that galaxies do form, we can then we can place an upper bound on $2 \leftrightarrow 2$ scatterings from galaxy mergers. Consider the behavior of the SIDM halo in a galaxy collision. Let $n_g$ be the number density of the SIDM particles in the halo. The mean free path of an SIDM particle in the halo is $1/n_g\sigma$; if this is shorter than the typical galactic halo radius $r_g \sim 10$ kpc, then most SIDM particles will undergo collisions in the process of galaxy merger. If this is the case, then we expect the galactic halos to "bounce" during a collision, while the baryons will not bounce, and some of the disk material may be ejected. Taking $r_g \sim 10$ kpc and $n_g \sim 10^4 \bar{n}$, we find that

$$1/\bar{n}\sigma > 100 \text{ Mpc} \tag{23}$$

is the condition that galaxy mergers not be significantly perturbed by SIDM scattering in the halo.

A second argument against the SIDM particles remaining in pressure equilibrium down to the present comes from peculiar motions of the galaxies themselves. If a galaxy has a typical peculiar velocity $v_g$, then the dark matter in the halo will feel a ram pressure force $F_{ram} \sim m'\bar{n}\sigma v_g^2$. The typical gravitational binding force on the halo dark matter is $F_g \sim m'v_c^2/r_g$, where $v_c$ is the SIDM velocity in the galaxy. The requirement that ram pressure not strip the halo off of the galaxy is $F_g > F_{ram}$, and if we take $v_g \approx v_c \approx 100$ km sec$^{-1}$, then this constraint is

$$1/\bar{n}\sigma > r_g \sim 10 \text{ kpc} \tag{24}$$

Note that our limits in equations (20), (22) and (23) are contradictory, while equations (20), (22) and (24) allow only a tiny window for the present-day mean free path of the SIDM particles. Neither of these arguments is ironclad. The entire process of galaxy formation, including mergers and collisions is not completely understood, and perhaps there is a way to incorporate "bouncing" halos into it. Our second argument assumes that the galaxy velocity relative to the dark matter background will be of order typical galaxy relative velocities. However, we would expect the galaxy to be moving with roughly the same velocity as the dark matter flow in its vicinity, decreasing this relative velocity substantially. However, in order to reduce this relative velocity to zero, the dark matter flow would have to be completely smooth on scales below the galaxy radius, which seems unlikely.

In either case, it is clear that we would have to substantially modify our picture of galaxy evolution if $2 \leftrightarrow 2$ scatterings were important down to the present, so we can consider this as another constraint on SIDM. In principle, we have no idea what the coupling strength of the $2 \leftrightarrow 2$ reactions is compared to the $3 \leftrightarrow 2$ reactions in the absence of a specific model; however, it is unreasonable to construct a model with strong $3 \leftrightarrow 2$ coupling and weak $2 \leftrightarrow 2$ reactions. If we write the $2 \leftrightarrow 2$ interactions in a form similar to eq. (6), the result is

$$\Gamma(2 \leftrightarrow 2) = \lambda_2 m' \left( \frac{n'}{m'^3} \right). \tag{25}$$

Using the fact that after decoupling the number density is related to the critical density by $n' = \rho_c/(m'a^3)$ one can derive the relation between $m'$ and $\lambda_2$ such that pressure equilibrium is maintained today; the result is

$$m' \lesssim 10^7 \lambda_2^{\frac{1}{3}} \text{ eV}. \tag{26}$$

If pressure equilibrium is not maintained down to the present, the inequality in eq. (26) must be violated. However, for the mass ranges considered here, this would require $\lambda_2 \ll \lambda$, which is very unnatural from a particle physics standpoint. This point is implicit in the CMH paper.

However, because it may be unrealistic from an astrophysical point of view to assume that pressure equilibrium is maintained down to the present day, we must consider what happens if the $2 \leftrightarrow 2$ interactions decouple at some earlier epoch. When the $2 \leftrightarrow 2$ interactions decouple, the SIDM particle may free-stream. While Jeans suppression freezes in the perturbation amplitude as it oscillates, free–streaming erases it.

To determine the largest possible effect which free-streaming can produce, we have considered the rather unrealistic "maximal free-streaming" scenario in which there are no $2 \leftrightarrow 2$ interactions, so pressure equilibrium is maintained solely by the $3 \leftrightarrow 2$ interactions. To properly account for free streaming, one must solve the relativistic Boltzmann equation. This calculation is discussed in detail in Schaefer & de Laix (1995); here we simply give the results. In Fig. 3, we show the difference between Jeans and free-streaming suppression for an SIDM model with $\lambda = 10^{-3}$ and $m' = 10$ keV. The solid curve is the transfer function for the SIDM model with pressure equilibrium maintained down the present, while the dashed curve is the transfer function for the maximal free-streaming model. We find that even with maximal free-streaming, the difference for the various rms fluctuations we have examined in this paper is insignificant. The reason is that the free-streaming length when the particle interactions decouple is roughly equal to the Jeans length at decoupling. Even without free-streaming, scales which lie below the Jeans length are strongly suppressed and do not contribute much to any of our mass fluctuation calculations. Hence, suppressing power on these length scales even further through free streaming does not affect our calculations. Our results for the mass fluctuation amplitude $\sigma$ on the various length scales we have examined differ at most by a few percent with or without free-streaming. This shows that the structure formation constraints derived here depend only on the strength of the $3 \leftrightarrow 2$ coupling and are independent of the strength of the $2 \leftrightarrow 2$ interactions.

## 5 Conclusions

It was originally hoped that the SIDM model would be able to circumvent the problems associated with the HDM and CDM models by producing a power spectrum intermediate between them. Unlike ordinary dark matter, the ability of SIDM to reheat itself means that it will have a much higher temperature during structure formation than corresponding CDM with the same mass. Preliminary investigation by CMH and by Machacek (1994) indicated that there might be viable SIDM models as long as decoupling occurred before the particle dominated the expansion of the universe. We have done a more extensive analysis and found that in fact, no SIDM model can fit current observations. It is clear from Fig. 1 that no SIDM model can simultaneously satisfy the constraints from the IRAS galaxy survey and damped Ly$\alpha$ systems. With a normalization consistent with COBE, SIDM models will either underproduce small-scale structure or overproduce large–scale structure. The problem is that SIDM relies on either Jeans or free streaming suppression to eliminate small-scale power. Regardless, the suppression is just too large to fit observation when it is significant. Furthermore, models which decouple early enough to lack significant Jeans suppression do not differ much from CDM models, which produce an excess of small-scale power when COBE normalized.

We have shown that structure formation constraints are virtually independent of the strength of the $2 \leftrightarrow 2$ couplings, but it is possible to constrain the $2 \leftrightarrow 2$ couplings using arguments about galaxy structure. Realistic particle physics models should have strong $2 \leftrightarrow 2$ couplings as well as $3 \leftrightarrow 2$ couplings, which implies that pressure equilibrium will be maintained to the present day. In such cases, during galaxy mergers, SIDM halos would interact and some of the disk material could be ejected. Also, the interaction of the halo with the background SIDM due to the galaxy peculiar velocity can strip the galaxy of its halo, another undesirable result. These arguments suggest that the $2 \leftrightarrow 2$ couplings may have to be unnaturally small.

Our work adds new constraints to the SIDM model. We have shown that regardless of the strength of $2 \leftrightarrow 2$ couplings, observations of galaxies and damped Lyman–$\alpha$ systems probably rule out SIDM. We have also been able to constrain SIDM based upon the $2 \leftrightarrow 2$ coupling strength and its effects on galaxy halos. Taken as a whole, SIDM does not appear to be a viable model for structure formation.

We thank Andy Gould for helpful discussions. A.A. de L. and R.J.S. were supported by the Department of Energy (DE-AC02-76ER01545). R.K.S. was supported by NASA (NRA-91-OSSA-11 and NRA-93-OSS-05).

# Figure Captions

**Fig. 1** The range of values for the SIDM particle mass, $m'$, and number-changing coupling $\lambda$, showing the limits for our two structure formation constraints, for the case where $2 \leftrightarrow 2$ interactions maintain the pressure equilibrium down to the present. Masses above the solid line in the horizontally hatched region produce excessive structure on $8h^{-1}$ Mpc scales. Masses below the dot–dashed line in the vertically hatched region produce insufficient early collapse for making damped Lyman–$\alpha$ systems.

**Fig. 2** The solid curves show the SIDM transfer functions for a coupling of $\lambda = 10^{-3}$ and masses $m' = 1$, 10 and 100 keV (no free–streaming). Also shown are the HDM (dotted), CDM (dashed) and mixed dark matter (20% HDM, 80% CDM) (dot–dashed) transfer functions.

**Fig. 3** A comparison of SIDM models ($\lambda = 10^{-3}$ and $m' = 10$ keV) with maximal free–streaming (dashed), i. e. no $2 \leftrightarrow 2$ interactions, and no free–streaming (solid).

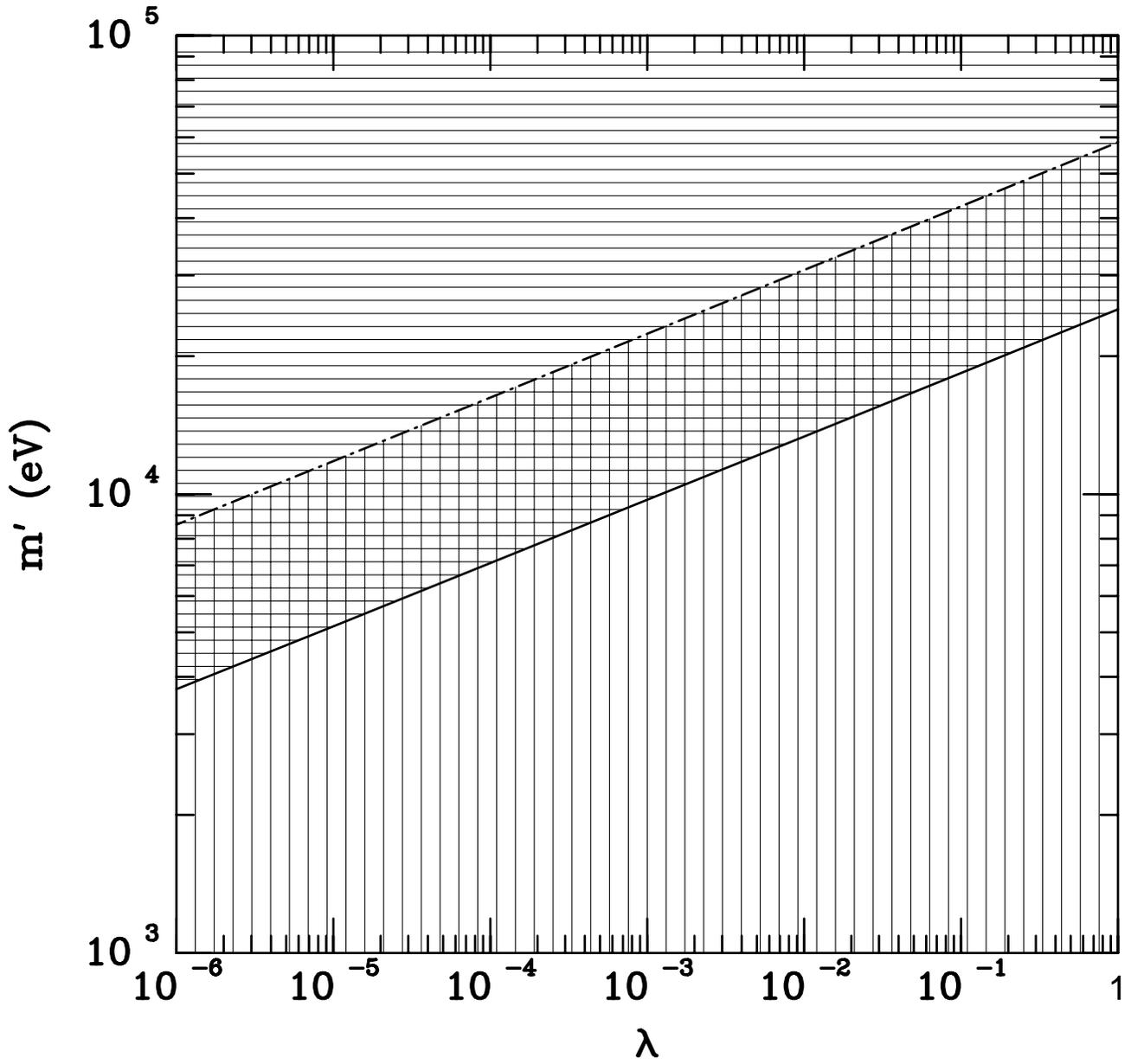

Figure 1. Ruled Out SIDM Parameter Space

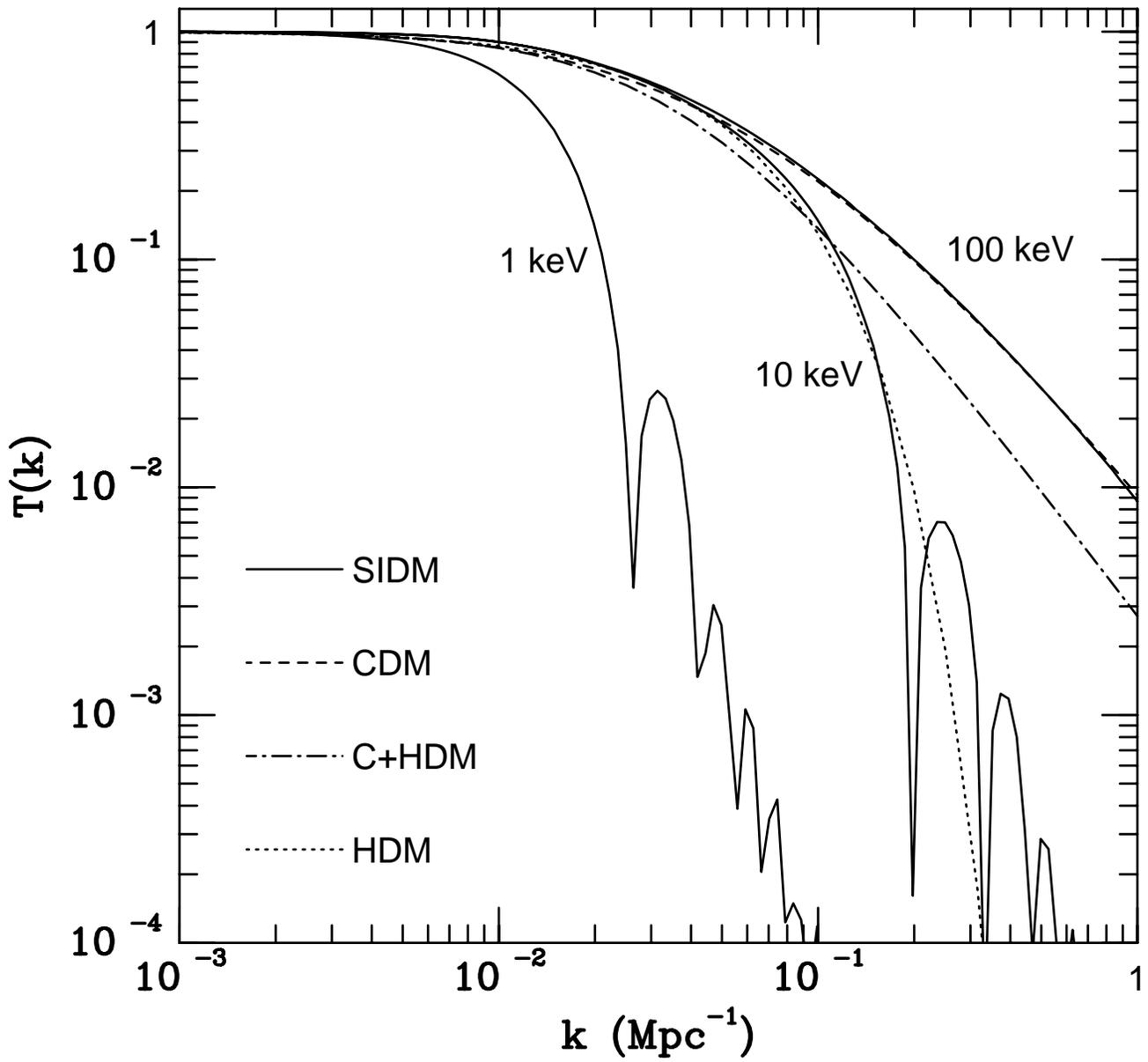

Figure 2. Transfer Functions in Different Models

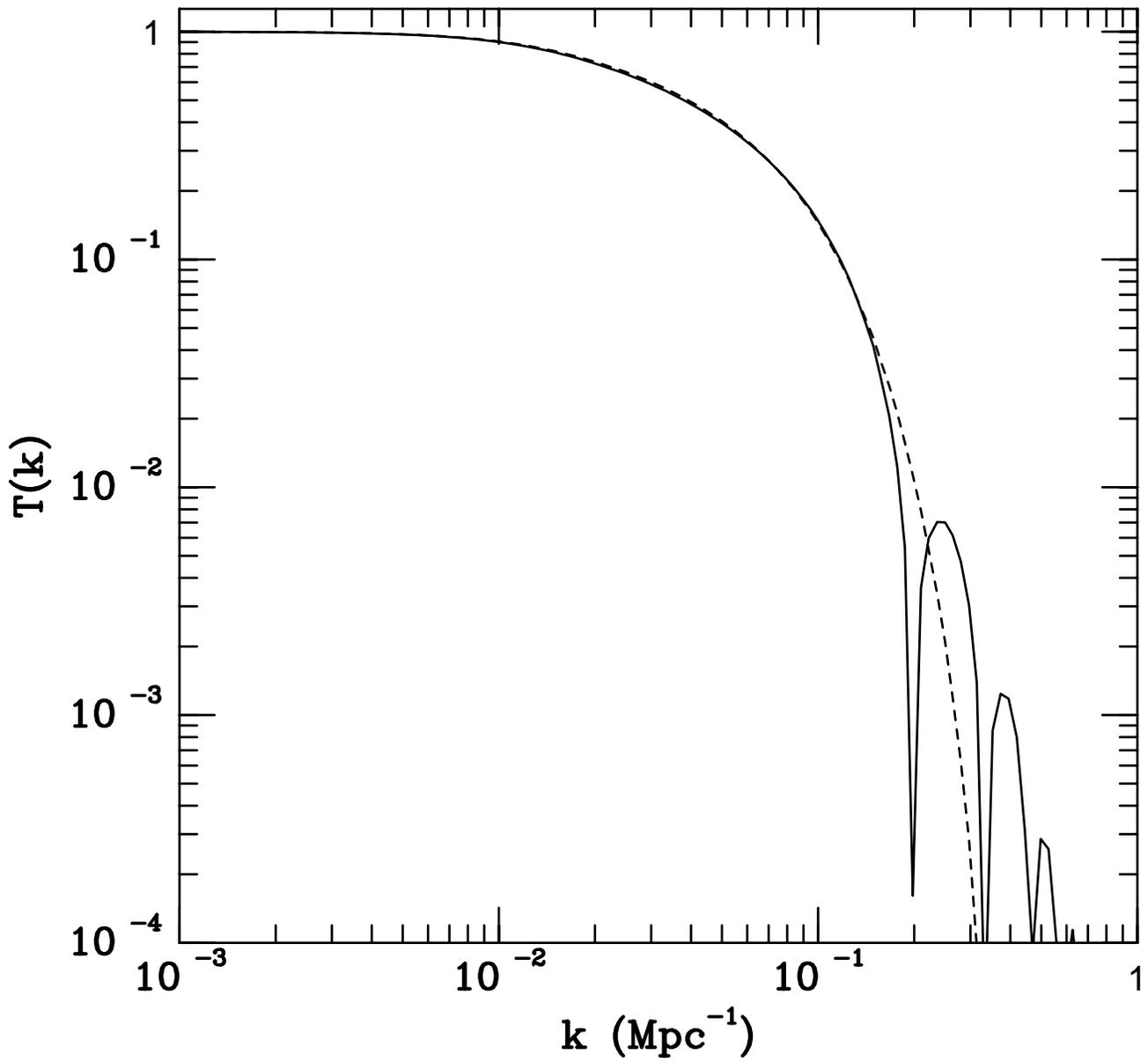

Figure 3. SIDM with/without Free Streaming